\documentclass{article}
\usepackage[utf8]{inputenc}
\usepackage[T1]{fontenc}

\usepackage{amsmath,amsthm,verbatim,amssymb,amsfonts,amscd, graphicx,mathrsfs,mathabx,stmaryrd,xcolor,algorithm,algorithmic}
\usepackage{titlesec}
\usepackage{graphics}
\usepackage{units}
\usepackage{hyperref}
\usepackage{enumerate}
\usepackage{moreverb}
\usepackage{stmaryrd} 
\usepackage{dsfont}
\usepackage{listings}
\usepackage{xcolor}
\usepackage{etoolbox}
\usepackage{framed}
\usepackage{lipsum}

\definecolor{codegreen}{rgb}{0,0.6,0}
\definecolor{codegray}{rgb}{0.5,0.5,0.5}
\definecolor{codepurple}{rgb}{0.58,0,0.82}
\definecolor{backcolour}{rgb}{0.95,0.95,0.92}

\lstdefinestyle{mystyle}{
    backgroundcolor=\color{backcolour},   
    commentstyle=\color{codegreen},
    keywordstyle=\color{magenta},
    numberstyle=\tiny\color{codegray},
    stringstyle=\color{codepurple},
    basicstyle=\ttfamily\footnotesize,
    breakatwhitespace=false,         
    breaklines=true,                 
    captionpos=b,                    
    keepspaces=true,                 
    numbers=left,                    
    numbersep=5pt,                  
    showspaces=false,                
    showstringspaces=false,
    showtabs=false,                  
    tabsize=2
}

\lstdefinestyle{mystyle}{
    backgroundcolor=\color{backcolour},   
    commentstyle=\color{codegreen},
    keywordstyle=\color{magenta},
    numberstyle=\tiny\color{codegray},
    stringstyle=\color{codepurple},
    basicstyle=\ttfamily\footnotesize,
    breakatwhitespace=false,         
    breaklines=true,                 
    captionpos=b,                    
    keepspaces=true,                 
    numbers=left,                    
    numbersep=5pt,                  
    showspaces=false,                
    showstringspaces=false,
    showtabs=false,                  
    tabsize=2
}

\newcommand\enclosebox[2]{%
  \BeforeBeginEnvironment{#1}{\begin{#2}}%
  \AfterEndEnvironment{#1}{\end{#2}}%
}

\newtheorem{theorem}{Theorem}
\theoremstyle{definition}

\theoremstyle{remark}

\enclosebox{theorem}{framed}
\enclosebox{example}{shaded}
\enclosebox{proof}{leftbar}

\begin{document}

\title{\Huge{Approximating Optimal Asset Allocations \\ using Simulated Bifurcation}}
\author{Thomas \textsc{Bouquet} \\ Mehdi \textsc{Hmyene} \\ François \textsc{Porcher} \\ Lorenzo \textsc{Pugliese} \\ Jad \textsc{Zeroual}}
\date{May 2021}
\maketitle

\vspace{80pt}

\begin{center}
\includegraphics[height=130px, width=130px]{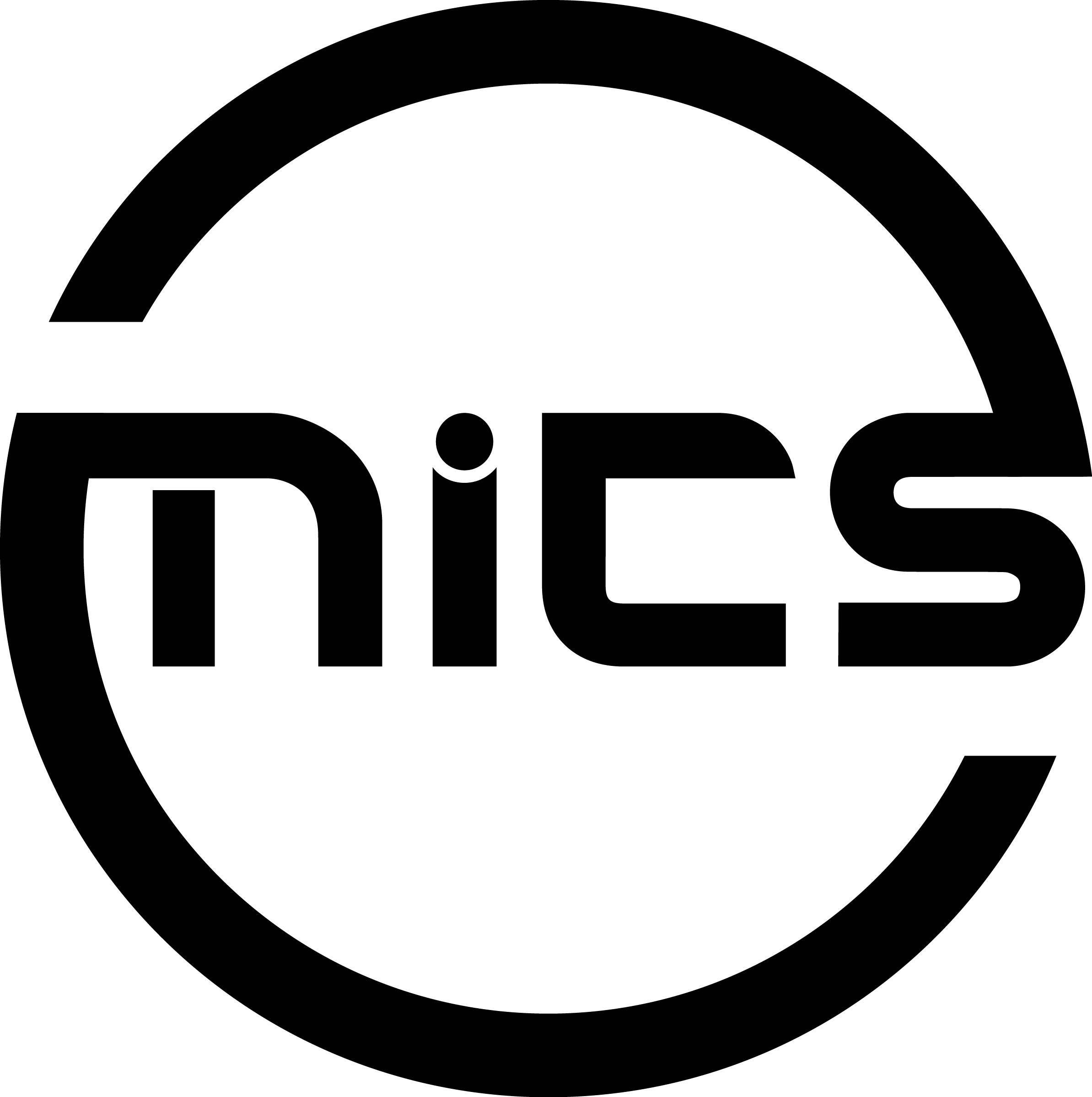} 
\end{center}

\begin{center}
\includegraphics[height=120px, width=400px]{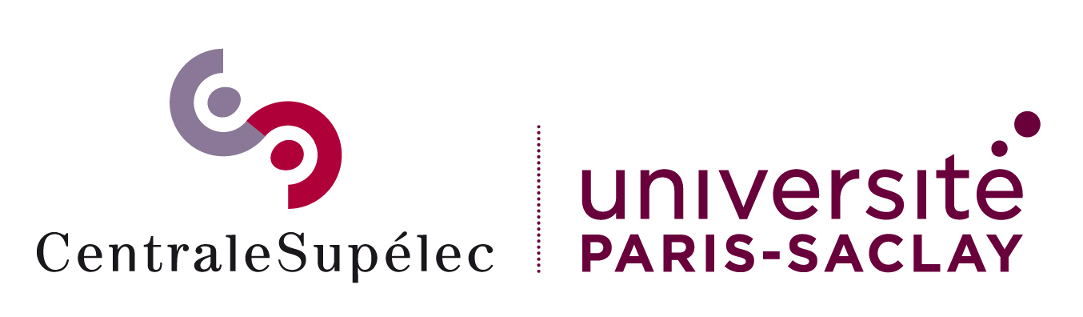}
\end{center}

\newpage

\Large
 \begin{center}
Abstract
\end{center}

\hspace{10pt}

\normalsize

This paper investigates the application of \textit{Simulated Bifurcation} algorithms to approximate optimal asset allocations. It will provide the reader with an explanation of the physical principles underlying the method and a \textsc{Python} implementation of the latter applied to 441 assets belonging to the S\&P500 index. In addition, the paper tackles the problem of the selection of an optimal sub-allocation; in this particular case, we find an adequate solution in an unrivaled timescale.

\vspace{50pt}

\Large
 \begin{center}
Acknowledgments
\end{center}

\hspace{10pt}

\normalsize

We would like to thank Damien \textsc{Challet} (Full professor HDR; Associate editor for \textit{Quantitative Finance}, \textit{Journal of Economic Interaction and Coordination}, \textit{Journal of Statistical Mechanics: theory and experiments}; Co-chief editor of \textit{Market Microstructure and Liquidity}) for his many useful pieces of advice and his guidance throughout the year. All remaining errors are ours. \\

We also thank Romain \textsc{Perchet} (Head of Multi-Asset Team for Quant Research Group at \textsc{BNP Paribas Asset Management}) for introducing us to the field of portfolio optimization theory and for providing us with resourceful data.\\

Finally, we thank the members of the MICS (\textsc{Research laboratory in Mathematics and Computer Science at CentraleSupélec}) and \textsc{CentraleSupélec} (Graduate School of Engineering of the \textsc{Paris-Saclay University}) for giving us the opportunity of undertaking research which falls under the scope of our general engineering curriculum.

\newpage

\tableofcontents

\newpage

\maketitle

\section{Introduction}

Efficient resolution of integer optimization problems is a major computational challenge, as such formulations can be found in almost any field that uses computer-aided calculations, from operations research to quantitative finance. That being said, and despite the significant progress made in the last decades, most existing algorithms have proven impractical to apply when the considered datasets exceed one hundred elements due to the exponential time complexities deriving from the combinatorial nature of the problems. \\

\vspace{0.3cm}

\hspace{-0.5cm}\begin{minipage}{0.45\textwidth}
Rather than directly seeking optimal configurations, a new method, commonly known as \textit{Simulated Bifurcation} and strongly inspired by the field of quantum physics, focus on obtaining explicit solutions to simpler problems and then slightly modifying such candidates in order to converge towards an approximated solution of the initial problem. The interest of such a process lies in its temporal efficiency: indeed, recent simulations have shown that such an approach can significantly shorten the execution time.

\end{minipage} \hfill
\begin{minipage}{0.5\textwidth}
\vspace{-0.8cm}
\begin{figure}[H]
\centering
\includegraphics[scale=0.65]{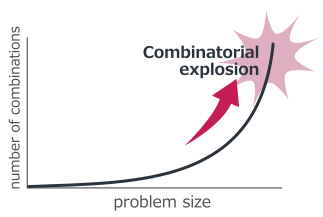}
\caption{Illustration from \textsc{Toshiba} SBM}
\end{figure}
\end{minipage}

\vspace{0.7cm}

In view of such promising results, he objective of our study will be to analyze and apply this class of algorithms to the fundamental problem of optimal asset allocation (aiming at maximizing risk-adjusted returns over a given time horizon), drawing on recent sources such as \cite{alpaca} and \cite{combinatorial}.

\vspace{0.5cm}

\section{State of the art}

   \subsection{\textit{NP-hard} optimization}
    
Many problems in fields such as computer science, economics or social sciences are classified as \textit{NP-hard}, which means that they are at least as difficult to solve than problems in NP (set of non-deterministic polynomial acceptable problems, for which a polynomial solution does not exist as far as we assume that $P \neq NP$); graph coloring, the clique problem or \textit{Knapsack} are some well-known examples belonging to the \textit{NP-hard} set. \\

As mentioned in the introduction, the number of possible solutions increases exponentially with the size of the problem data, making \textit{NP-hard problems} impractical to solve with a brute-force algorithm (which enumerates all possible configurations and elects the very best one) as soon as we consider more than a dozen elements. In response to this combinatorial explosion, methods yielding exact results, such as the famous concept of dynamic programming introduced in the 50s by \textsc{Richard Bellman}, are based on a subdivision of the initial complex problems into several easier to solve sub-problems, whose solutions are combined in order to get to the global solution. The simplex algorithm for linear problems and the Branch \& Bound algorithm are other leading references when it comes to the accurate resolution of \textit{NP-hard} problems. \\

On the other hand, one might only be interested in obtaining an approximation of the global optimum it if means that this output can be computed in only a fraction of the computation time. In these cases, another set of algorithms can be used including, among others, methods based on heuristics / metaheuristics (for instance, the famous genetic algorithm uses an heuristic inspired by the theory of natural evolution). This paper will focus on \textit{Simulated Bifurcation}, an algorithm that belongs to this type of methods and that has many interesting properties.
    
\subsection{Portfolio optimization}
    
Portfolio optimization with discrete weights belongs to the large family of combinatorial optimization problems.
Combinatorial optimization is a type of optimization that combines various techniques derived from discrete mathematics and computer science in order to find the best solution within a set of feasible tuples. The main challenge associated with these problems is to find an optimal solution in a reasonable execution time. Indeed, most instances of combinatorial optimization belong to the class of \textit{NP-hard} problems, for which efficient resolution algorithms rarely exist. It is thus important to find numerical methods that overcome this difficulty, even if that means obtaining an approximate solution. \\

Optimizing the risk-adjusted return of a financial portfolio when we limit the scope of our study to discrete weights is precisely such a \textit{NP-hard} problem (the complete mathematical framework, introduced by \textsc{Harry Markowitz} in \cite{original}, is presented in the following section). One of the most promising approaches to perform portfolio optimization when the number of assets exceeds 100 is called \textit{Simulated Bifurcation} and is based on a parallel with quantum physics. The first cornerstone was laid in \cite{NP-complete}, an article dating from 1998 and dealing with non-linear optimization problems (the example of the futures market, characterized by a non-linear constraint connecting the investor's wealth and the margin requirements of each underlying asset, received special attention). The authors show how the application of quantum theory, combined with stochastic calculus, makes it possible to determine a set of optimal portfolios for a given risk. In fact, the solutions found depend on the correlation matrix of the assets, which is taken to be stochastic under certain conditions, inspired by the original idea of \textsc{Wigner} and \textsc{Dyson} of replacing the Hamiltonian of a complex deterministic system by a random matrix. \\

In later years, researchers realized that the analogy could be extended to use quantum physics in the resolution of combinatorial problems; the fundamental principle of such a reasoning is the fact that an integer can be represented as a sequence of spins thanks to its binary representation. As a result, very recent papers such as \cite{combinatorial} and \cite{alpaca} have demonstrated the merits of this approach, in particular as regards the computing speed, by simulating adiabatic evolutions of classical nonlinear Hamiltonian systems (the adiabatic nature of the process allowing them to closely monitor the evolution of the solution). \\

\textit{\textsc{Ising} machines} designed to find the fundamental state of systems having \textsc{Ising}-like energies have also attracted a lot of attention recently because of the great number of combinatorial problems that can be easily mapped to the general form required for \textit{Simulated Bifurcation} to be applicable (\textit{Max-Cut} being a well-known instance of a problem that can be reformulated). A telling example is given by the authors of \cite{alpaca}, who have managed to solve the optimal trading trajectory problem with an unprecedented computational speed. \\

\newpage

\section{The \textsc{Markowitz} model}

The \textsc{Markowitz} representation (proposed by Harry \textsc{Markowitz} in \cite{original}) provides us with a concise and pleasant mathematical framework to manipulate financial portfolios, which justifies the fact that this standard will be adopted hereafter. \\

In order to simplify the formulation of the optimization problem, the key idea is to describe the set of feasible portfolios (i.e. a set of portfolios subject to certain assumptions) and then to choose an \textit{optimal portfolio} within that set. To that end, we represent the portfolio as a \textbf{linear combination of assets} within the considered collection. We can then define two fundamental characteristics:

\begin{itemize}
    \item \textbf{The expected return $\mu$}, which corresponds to the mathematical expectation. For a portfolio $P$ with $N$ assets $(A_i)_{1\leq i \leq N}$, we have $\mu(P) = \sum\limits_{i} w_{i} A_{i}$ where the $w_i$ are the chosen weights;
    \item \textbf{The volatility $\sigma_P$}, defined thanks to the variance $\sigma_{P}^{2} = \sum\limits_{i,j} w_{i} w_{j} \sigma_{ij}$ where $\sigma_{ij}$ represents the covariance coefficient between assets $A_i$ and $A_j$.
\end{itemize}

Thus, if a portfolio has an expected return of $10\%$, the return we would obtain would be within a range defined by its volatility. For a volatility of $8\%$, we can expect a return between 2 and $18\%$, while for a volatility of $12\%$, we will have a return between -2 and $22\%$. This obviously emphasizes the importance of risk management and risk-adjusted returns: indeed, the second option might seem more appealing as the maximum possible return is greater but having a negative return is just as likely, whereas the first option almost surely implies a positive return. We therefore introduce a parameter $\gamma$ to represent the risk aversion of the market players. \\

Finally, the acquisition of assets is not free (brokerage fees, account and transaction fees, taxes on financial transactions) and we must take into account such costs whenever a purchase is made. \\

\begin{figure}[!h]
\centering
\includegraphics[scale=0.4]{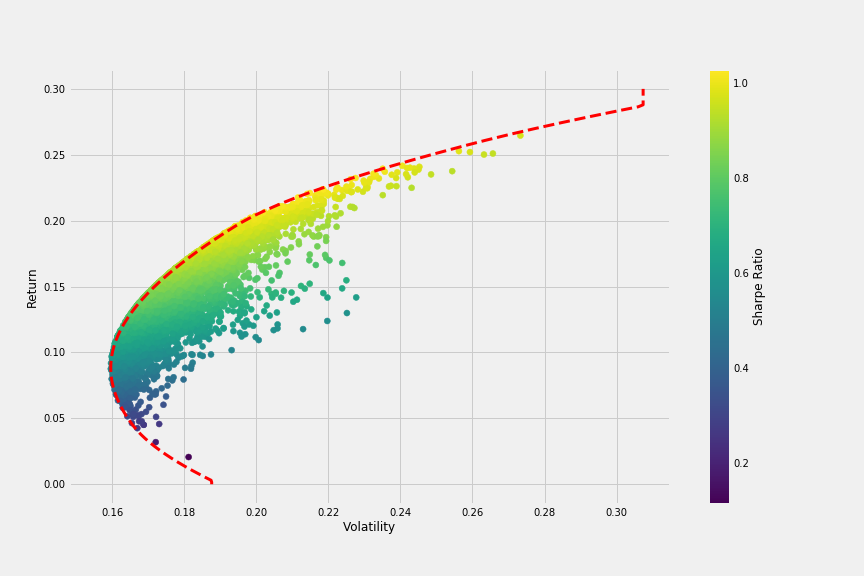}
\caption{Example of \textsc{Markowitz} efficient frontier (set of portfolios that maximize the expected return for a given risk), illustration from \cite{markowitz}}
\end{figure}

\newpage

\section{Formulation of the constrained optimization problem}

With such a framework, we can reformulate the portfolio optimization problem. For a collection of $N$ assets, we define:

\begin{enumerate}
    \item $w_t = \begin{bmatrix}
w_1\\
\vdots\\
w_N
\end{bmatrix}$ the weighting vector. It accounts for the capital invested on asset $i$ by the agent studied at time $t$;
\item $\mu_t$ the return vector which provides the expected value of each asset at time $t$;
\item $\Sigma_t$ the covariance matrix at time $t$ (for the sake of simplicity, will assume that $t \mapsto \Sigma_t$ is constant);
\item $\Lambda_t$ the matrix modelling the costs of acquisition of the assets at time $t$. 
\item For the sake of clarity, we also define $\Delta w_t = w_t - w_{t-1}$.
\end{enumerate}

In view of the above, the utility function to be maximized is:
$$\sum\limits_{t} \left [ w_{t}^{T}\mu_{t} - \frac{\gamma}{2} w_{t}^{T}\Sigma_{t} w_{t} - \Delta w_{t}^{T} \Lambda_{t} \Delta w_{t} \right ]$$
This expression is composed of three key terms:

\begin{itemize}
    \item $w_{t}^{T} \mu_{t}$ represents the expected return: indeed, $w_{t}^{T} \mu_{t} = \sum\limits_{i} w_{i} \mu_{i}$;
    \item $\gamma (-\frac{1}{2} w_{t}^{T} \Sigma_{t} w_{t})$ represents the risk induced by the volatility of the portfolio, as $w_{t}^{T} \Sigma_{t} w_{t} = \sum\limits_{i,j} w_{i} w_{j} \sigma_{ij} = \sigma_{P}^{2}$. As regards the negative sign, it seems natural that diversification is key to finding an optimal portfolio. Indeed, let us consider the example of two assets $A$ and $B$ with $P = w.A + (1-w).B$ the portfolio. If the companies are in the same sector, i.e. $Cov(A,B) > 0$, then a decrease in the value of one asset will result in a decrease in the value of the other. Thus, if the sector loses momentum, the impact on the portfolio will be all the more disastrous. If $Cov(A,B) < 0$, then $\sigma_{P}^{2} = w^{2}.\sigma_{A}^{2} + (1-w)^{2}.\sigma_{B}^{2}$. This reduces the volatility of $P$ and therefore limits the risk;

    \item The last term simply expresses the sum of the costs caused by the change in the quantity of assets owned (brokerage fees).
\end{itemize}

\vspace{.1cm}

\subsection{Simplifying assumptions}

In this section, we will present simplifying assumptions that will allow us to draw an analogy with quantum physics (ground state of a $\varphi(n)$-particles system with $\varphi$ linear with respect to $n$).

\subsubsection{Integer weighting}

It is assumed that $\forall i \leq N, w_i \in \mathbb{N}$. This may seem surprising as one would expect to have $w_i \in [0,1]$ to represent the fraction of capital allocated to an asset. That being said, if we have a capital $C$, considering $\frac{w_i}{C}$ with $w_i \in \mathbb{N}$ represents the part of the budget allocated to the asset $i$. Indeed, we have \textit{a priori} $w_i \in \mathbb{R}^{+}$ because the weights are necessarily positive. Restricting the problem to integers provides us with a straightforward binary representation of the vector $w$. This representation would also be exist on the set of positive real numbers since \textit{floats} are encoded in a binary way, but the transition from one basis to the other requires more precaution, while the theoretical foundation of the method remains unchanged.

\subsubsection{Budget constraints}\label{budget}

The previous constraint invites us to define a time-independent capital $C > 0$. We therefore suppose that $\forall t, \sum\limits_{i} w_{i} \leq C$. This assumption is quite natural, as it is clear that if we do not impose a limit on the quantity of assets one can own, the optimal solution would be unbounded. Furthermore, this hypothesis accurately transcribes the reality of the market where only a finite number of assets can be acquired. One could also consider a case where $C(t)$ would depend on time (a trader could be given a certain number of units per day, for instance), but such a case will not be discussed in this paper. \\

An even stronger hypothesis can be formulated to avoid pathological cases in which the agent would invest all his capital into a single asset (thus defying the principle of diversification). The uniform constraint $\forall t, \forall i, 0 \leq w_i \leq C/N$ makes it possible to force diversification but has several drawbacks. We can relax this constraint by noticing that:

\begin{enumerate}
    \item The capital $C$ is arbitrary. Indeed, whatever its value, by renormalizing the $w_i$ we can obtain a different capital. We can therefore choose as large a capital as we wish;
    \item The results are more meaningful when the budget is large. Intuitively, one could say that setting a higher initial capital allows the algorithm to shuffle a larger number of integers and therefore to refine the precision of the selected weights. We will see in the following sections that this intuition is verified;
    \item Deriving from the representation of $w_i$ as binary numbers, we know that there exists $\alpha\in\mathbb{N}$ such that $\forall i, w_i \leq 2^\alpha -1$. \\
\end{enumerate}

It is thus possible to modify the constraint on $w_i$ by defining a constraint on $\alpha$. An easy way to guarantee that $\forall t, \sum\limits_{i} w_{i} \leq C$ is to make sure that $(2^\alpha -1)N \leq C$ for example. \\

That being said, such an assumption is not mandatory. Indeed, as evoked above, $C$ can take arbitrarily large values, so one could choose $\alpha$ arbitrarily large as well. As a conclusion, the choice of $\alpha$ is a trade-off between overall computation time and accuracy of the approximated solution. 

\subsubsection{Transactional costs}

As regards the determination of the $\Lambda_{t}$ matrix for a given $t$, several approaches are possible. In a simple case where transactional fees are constant and time-independent, we have that the cost can be expressed as $\forall t, \forall i, c_{i}(t,t+1) = c$ where $c$ is a constant.
We can thus rewrite the last term of the problem $\Delta w^{T} \Lambda \Delta w$ as $c ||\Delta w||^2$. Indeed, if we suppose that we can only change the quantity of assets owned by one unit between $t$ and $t+1$ (this is the case in the common stock market), $|| \Delta w_i||^{2} = \sum\limits_{i}\epsilon_{i}^{2}$ where $\epsilon_{i} = w_{it} - w_{i(t+1)}$ and $||\Delta w_{i}||^{2} = w_{it} - w_{i(t+1)}$. \\

For the sake of simplicity, transactional and brokerage fees are going to be neglected hereafter (i.e. $\Lambda_t = 0_{\mathcal{M}_n}$); in addition to simplifying the model, such an approximation will prove necessary in order to draw a parallel between the problem described here and the equivalent quantum system.

\subsubsection{Time-independence}

Let us notice that it is be reasonable  to consider that the functions $t \mapsto \Sigma_t$ and $t \mapsto \mu_t$ are constant if large fluctuations on the market are not likely to occur over the investment horizon (for instance, a change of correlation between the stock price of two assets is quite an extraordinary event).  Such an assumption makes each term of the sum in the expression of the utility function time-independent if we neglect the transactional costs.

\subsection{Spin representation}

In summary, we can now get rid of the $t$ and simply write the weighting vector $w$. Besides, we know that for all $i \in \llbracket 1,N \rrbracket, \; w_i \in \mathbb{N}$ and $w_i \leq C/N$. Therefore, by choosing $\alpha = \lfloor \log_2 (C/N) \rfloor + 1$ we can assert that $w \in \llbracket 0, 2^\alpha-1 \rrbracket^{N}$. Hence, the following property holds:

\begin{theorem}[Spin representation]
Let $U_{N\alpha} \in \mathbb{R}^{N\alpha}$ the vector filled with ones and $M_{N, \alpha} \in \mathcal{M}_{N\alpha, N}(\mathbb{R})$ be the matrix defined as follow:
$$\forall k \in \llbracket 1, N \rrbracket, \forall l \in \llbracket 1, \alpha \rrbracket, \;\; \left [M_{N, \alpha} \right ]_{kl} = \left\{
    \begin{array}{ll}
        2^{(k-1) - (l-1)\alpha } & \mbox{if } (k-1) - (l-1)\alpha \in \llbracket 0, \alpha -1 \rrbracket \\
        0 & \mbox{else} 
    \end{array}
\right.$$
Then,
$$\forall w \in \llbracket 0, 2^\alpha-1 \rrbracket^{N}, \; \exists ! \; s \in \lbrace -1, 1 \rbrace ^{N\alpha}, \; w = \frac{1}{2} M_{N, \alpha}^T \left (s + U_{N\alpha} \right )$$ $s$ is called the \textbf{spin representation} of $w$.
\end{theorem}

\begin{proof}

Let $w \in \llbracket 0, 2^\alpha-1 \rrbracket^{N}$.\\\\ Let $i \in \llbracket 1, N \rrbracket$. Let us notice that:
$$\exists ! \; (b_{0,i}, b_{1,i}, ..., b_{\alpha-1,i}) \in \llbracket 0,1 \rrbracket ^\alpha, \; w_i = \sum_{k = 0}^{\alpha - 1} 2^k b_{k,i} \;\;\; (\mbox{binary representation of } w_i)$$
\\
We can then introduce the vector $b \in \llbracket 0,1 \rrbracket ^{N\alpha}$ such that: $b = (b_{0,1}, ..., b_{\alpha-1,1}, b_{0,2}, ..., b_{\alpha-1,N})^T$ which is unique by construction. The previous equation can then be rewritten in terms of $b$'s components: $$\forall i \in \llbracket 1,N \rrbracket, \; \; w_i = \sum_{k=1}^{\alpha } 2^{k-1} b_{(i-1)\alpha + k}$$
\\
We can finally introduce the vector $s \in \lbrace -1,1 \rbrace ^{N\alpha}$ such that $s = 2b - U_{N\alpha}$ which is also \textbf{unique} by construction ($s \in \lbrace -1,1 \rbrace ^{N\alpha}$ since $\forall x \in\lbrace 0,1 \rbrace, \; 2x-1 \in \lbrace -1,1 \rbrace$).
\\\\
Let us denote $p = \frac{1}{2} M_{N, \alpha}^T \left (s + U_{N\alpha} \right ) = M^Tb\in \mathbb{R}^N$. Let $k \in \llbracket 1, N \rrbracket$. In the following, we will just write $M_{N, \alpha}$ as $M$ for the sake of clarity.

\begin{align*}
    p_k & = \left [ M^Tb \right ]_k \\
    & = \sum_{i=1}^{N\alpha} \left [ M^T \right ]_{ki}b_i \\
    & = \sum_{i=1}^{N\alpha} M_{ik}b_i \\
    & = \sum_{u=0}^{N-1} \sum_{v=1}^{\alpha } \underbrace{M_{u\alpha +v,k}}_{
        \begin{array}{ll}
            M_{u\alpha +v,k} \neq 0 & \Leftrightarrow u\alpha + v - (k-1)\alpha - 1 \in \llbracket 0, \alpha -1 \rrbracket \\
            & \Leftrightarrow (u-k+1)\alpha + \underbrace{v-1}_{\in \llbracket 0, \alpha -1 \rrbracket} \in \llbracket 0, \alpha -1 \rrbracket \\
            & \Leftrightarrow u = k - 1
        \end{array}
    }b_{u\alpha +v} \\
    & = \sum_{v=1}^{\alpha } M_{(k-1)\alpha +v,k}b_{(k-1)\alpha +v} \\
    & = \sum_{v=1}^{\alpha } 2^{v-1} b_{(k-1)\alpha +v} \\
    & = w_k
\end{align*}
Hence, we get that for all $w \in \llbracket 0, 2^\alpha-1 \rrbracket^{N}$, there is a unique vector $s \in \lbrace -1,1 \rbrace ^{N\alpha}$ such that: $$w = \frac{1}{2} M_{N, \alpha}^T \left (s + U_{N\alpha} \right )$$
\end{proof}

\subsection{Identification of the \textsc{Ising} model}

The \textsc{Ising} model is used in statistical physics to approximate the behavior of a group of interacting two-state particles. Such a model helps physicists understand the interactions in ferromagnetic substances, and has a well-known explicit solution. In the previous sections, we have reduced the original optimization problem to one where we are looking for the configuration that minimizes a certain function of a set of two-state variables. Looking for the ground state of a ferromagnetic magnet and finding the optimal portfolio allocation are thus two very similar problems in this framework. \\

Typically, the energy of an \textsc{Ising} magnet is expressed as: $E = -\frac{1}{2}\sum\limits_{i,j} J_{i,j}s_i s_j + \sum_{i} h_i s_i$ where $J_{i,j}$ models the interaction between the particles $i$ and $j$ and where $h$ represents the magnetic field that acts on the particles. This model generally assumes that $i$ and $j$ must be \textit{neighbors} or spatially adjacent, as interactions between distant particles can be neglected. As the notion of \textit{neighbors} is difficult to define in a financial context, we will adopt a slightly more general framework in which $\forall i, \forall j, J_{i,j} \neq 0$. \\

Let us introduce the following notation: 
$$\forall n \in \mathbb{N}^*, \forall J \in \mathcal{M}_{n}(\mathbb{R}), \forall h \in \mathbb{R}^n, \forall s \in \lbrace -1,1 \rbrace^n, \; E_{Ising}(J,h,s) = -\frac{1}{2}s^TJs + s^Th$$

Using the new writing of the weighting vector, we are able to identify an Ising problem strictly equivalent to our original Markowitz one as follows.

\begin{theorem}[Models equivalence]
Maximizing the utility function of the Markowitz model is equivalent to minimize the Ising energy of the Ising model with $$J = -\frac{\gamma}{2}M_{N, \alpha} \Sigma M_{N, \alpha}^T \; \mbox{ and } \; h = \frac{\gamma}{2}M_{N, \alpha} \Sigma M_{N, \alpha}^TU_{N\alpha} - M_{N, \alpha}\mu$$
\\
Besides, for all spin vector $s \in \{-1,1\} ^{N\alpha}$ that is a spin vector of this Ising model: $$w = \frac{1}{2} M^T (s + U) \; \in \underset{w \in \llbracket 0, 2^\alpha-1 \rrbracket^{N}}{\mbox{argmax }} w^T\mu - \frac{\gamma}{2}w^T \Sigma w$$
\end{theorem}

\begin{proof}
For the sake of clarity, we will denote $M_{N, \alpha}$ as $M$ and  $U_{N\alpha}$ as $U$ in the following.
\\\\
Let $s \in \lbrace -1,1 \rbrace^{N\alpha}$ and $w = w(s) = \frac{1}{2}M^T(s+U)$. We have:
\begin{align*}
    w^T\mu - \frac{\gamma}{2}w^T \Sigma w & = \frac{1}{2}(s + U)^TM\mu - \frac{\gamma}{2} \frac{1}{2}(s + U)^TM\Sigma M^T\frac{1}{2}(s + U) \\
    & = \frac{1}{2}s^TM\mu + \frac{1}{2}U^TM\mu - \frac{\gamma}{8} \left [s^TM\Sigma M^Ts + 2s^TM\Sigma M^TU + U^TM\Sigma M^TU \right] \\
    & = \frac{1}{2} \left [-\frac{\gamma}{4}s^TM\Sigma M^Ts + s^T \left ( M\mu -\frac{\gamma}{2}M\Sigma M^TU \right ) \right ] + \frac{1}{2}U^TM\mu - \frac{\gamma}{8}  U^TM\Sigma M^TU \\
    & = - \left ( \frac{1}{2} \left [\frac{\gamma}{4}s^TM\Sigma M^Ts + s^T \left ( \frac{\gamma}{2}M\Sigma M^TU -  M\mu \right ) \right ] + \underbrace{\frac{\gamma}{8}  U^TM\Sigma M^TU - \frac{1}{2}U^TM\mu}_{f(U) \; \in \; \mathbb{R}} \right)
\end{align*}
By introducing the matrix $J = -\frac{\gamma}{2}M \Sigma M^T$ and the vector $h = \frac{\gamma}{2}M \Sigma M^TU - M\mu$, the previous equality can be rewritten as:
$$w^T\mu - \frac{\gamma}{2}w^T \Sigma w = - \left [ \frac{1}{2} E_{Ising}(J,h,s) + f(U) \right ]$$
\\
Let us now consider that $s \in \underset{s \in \lbrace -1,1 \rbrace^{N\alpha}}{\mbox{argmin}} E_{Ising}(J,h,s)$. Since, $\frac{1}{2} > 0$ and $f(U) \in \mathbb{R}$, we have:
$$s \in \underset{s \in \lbrace -1,1 \rbrace^{N\alpha}}{\mbox{argmin}} \frac{1}{2}E_{Ising}(J,h,s)+f(U) \Leftrightarrow s \in \underset{s \in \lbrace -1,1 \rbrace^{N\alpha}}{\mbox{argmax}} - \left [ \frac{1}{2} E_{Ising}(J,h,s) + f(U) \right ]$$
\\
Similarly, let us introduce $w_* \in \underset{w \in \llbracket 0, 2^\alpha-1 \rrbracket^{N}}{\mbox{argmax}} w^T\mu - \frac{\gamma}{2}w^T \Sigma w$ with $s_*$ its spin representation, and suppose that $w_* \neq w$. This leads to: 
$$w_*^T\mu - \frac{\gamma}{2}w_*^T \Sigma w_* > w^T\mu - \frac{\gamma}{2}w^T \Sigma w \Longleftrightarrow - \left [ \frac{1}{2} E_{Ising}(J,h,s_*) + f(U) \right ] > - \left [ \frac{1}{2} E_{Ising}(J,h,s) + f(U) \right ]$$
which contradicts the definition of $s$. Then, $w \in \underset{w \in \llbracket 0, 2^\alpha-1 \rrbracket^{N}}{\mbox{argmax}} w^T\mu - \frac{\gamma}{2}w^T \Sigma w$.
\\\\
We can then conclude that maximizing the utility function of the Markowitz model is equivalent to find the ground state of the Ising model with $J = -\frac{\gamma}{2}M \Sigma M^T$ and $h = \frac{\gamma}{2}M \Sigma M^TU - M\mu$ and that for all $s \in \underset{s \in \lbrace -1,1 \rbrace^{N\alpha}}{\mbox{argmin }} \left [\frac{\gamma}{4}s^TM\Sigma M^Ts + s^T \left ( \frac{\gamma}{2}M\Sigma M^TU -  M\mu \right ) \right ]$ :
$$\frac{1}{2} M^T (s + U) \; \in \underset{w \in \llbracket 0, 2^\alpha-1 \rrbracket^{N}}{\mbox{argmax }} w^T\mu - \frac{\gamma}{2}w^T \Sigma w$$
\end{proof}

\newpage

\section{Physical study of the quantum system}

It is common practice to draw parallels between different scientific disciplines to better understand the intricacies of a given problem. In that spirit, this section will focus on the study a quantum system that provides us with an approximate solution to a problem resembling \textsc{Ising}'s energy minimization.

\vspace{10pt}

\subsection{Understanding the physical origin of the Hamiltonian}

\subsubsection{\textsc{Josephson} junction and nonlinear resonator}

The physical system at the origin of the model studied hereafter is that of the resonator (supposedly ideal, i.e. made of superconductors) rendered non-linear by the presence of a \textsc{Josephson} junction (an illustration of which is presented below). The purpose of this subsection is to derive the Hamiltonian of such a system, necessary first step to understand the physical origin of the model introduced in the article. \\

\begin{figure}[!h]
\centering
\includegraphics[scale=0.5]{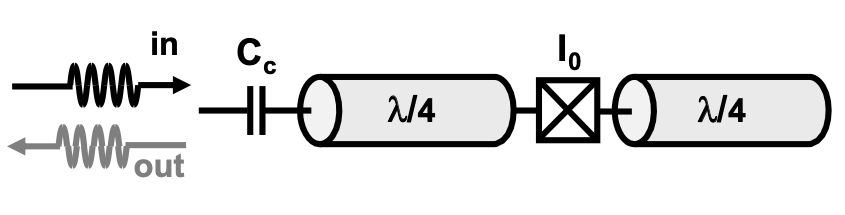}
\caption{Scheme of a nonlinear resonator (a \textsc{Josephson} junction of critical current $I_0$ is embedded in the middle of a $\frac{\lambda}{2}$ resonator; it is coupled to a $50\Omega$ transmission line through a coupling capacitor $C_c$ and probed in reflection by a microwave field), illustration from \cite{resonator}}
\end{figure}

\vspace{0.3cm}

The advantage of such a model lies in its easy conversion to an electrical system, on which it is undoubtedly easier to have a physical intuition. An equivalent circuit is presented below (the only unusual component is the \textsc{Josephson} junction whose fundamental property will be discussed in the following) and the values of the parameters presented can be explained using the basic properties of the circuit ($w_1 = 1/\sqrt{L_eC_e}$ being the resonance frequency of the junctionless circuit, $Z_0 \simeq 50 \Omega$, $L_e = \frac{\pi Z_0}{2w_1}$ and $C_e=\frac{2}{\pi Z_0 w_1}$). \\

\begin{figure}[!h]
\centering
\includegraphics[scale=0.45]{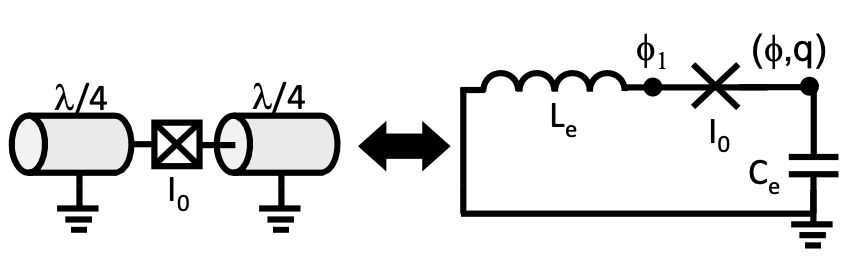}
\caption{Equivalence between the distributed nonlinear resonator and a series combination of an equivalent inductance $L_e$, capacitance $C_e$ and \textsc{Josephson} junction of critical current $I_0$, illustration from \cite{resonator}}
\end{figure}

\newpage

\begin{figure}[!h]
\centering
\includegraphics[scale=0.45]{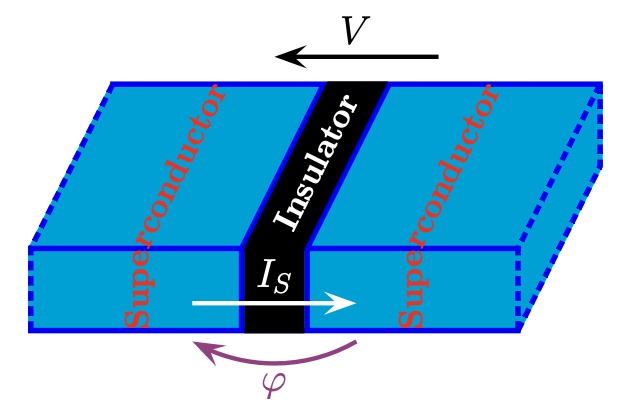}
\caption{A schematic representation of a superconducting \textsc{Josephson} junction (two superconducting electrodes are connected to each other via an insulating layer), illustration from \cite{josephson}}
\end{figure}

\vspace{10pt}

The \textsc{Josephson} junction obeys the differential equation $\dot \varphi = \frac{2e}{\hbar}V$ (the variables are defined in the illustrative figure above), thus defining an energy related to this junction of $\mathcal{H}_J = -E_J cos(\varphi)$ ($E_J$ being a constant related to the critical intensity). The other components of the Hamiltonian involve the capacitor (classically $\mathcal{H}_C = \frac{Q^2}{2C_e}$ with $Q$ the charge at the junction) and the coil (classically $\mathcal{H}_L = \frac{\phi_{1}^2}{2L_e}$ with $\phi_1$ magnetic flux). This leads to a global Hamiltonian of the system defined by:

$$\mathcal{H} = \mathcal{H}_J + \mathcal{H}_C + \mathcal{H}_L$$

As $I = \frac{\phi_1}{L_e} = I_0 \mathrm{sin}(\frac{\phi-\phi_1}{\phi_0})$ with $\phi_0 = \frac{\hbar}{2e}$ (same current flowing through the inductance and the junction), we can eliminate $\phi_1$, which makes it possible to perform a \textsc{Taylor} expansion of the Hamiltonian as a function of $\phi$ (the system operates far from the critical $E_J$ energy so the phase difference is negligible). An expansion to the fifth order provides:

$$\mathcal{H} = \mathcal{H}_C + \frac{\phi^2}{2L_t} - \frac{L_{j}^3 \phi^4}{24L_t \phi_{0}^2 L_{e}^3} + o(\phi^5)$$ with $L_t$ the sum of the inductances of the $L_j$ junction and $L_e$.  \\

Moreover, it turns out that it is possible to express $\phi$ and $Q$ as linear combinations of creation and annihilation operators ($(Q,\phi)$ is a pair of canonically conjugated variables like $(x,p)$). In order to understand this principle, let us consider a junctionless quantum LC circuit: we have $\mathcal{H} = \frac{Q^2}{2C_e} + \frac{\phi^2}{2L_e}$ for this circuit and $\mathcal{H} = \frac{p^2}{2m} + \frac{m w^2 x^2}{2}$ for a harmonic oscillator, which justifies an analogous definition of $a$ and $a^{\dagger}$. We then have the formulas $\phi = i \sqrt{\hbar Z_e/2} (a-a^{\dagger})$ and $Q = i \sqrt{\hbar /2Z_e} (a+a^{\dagger})$ with $Z_e = \sqrt{L_t/C_e}$ by generalizing the reasoning to the LC circuit with the junction (the resonance frequency is $w_r = 1/\sqrt{L_t C_e}$ as we need to take into account the inductance of the junction). \\

In order to derive the equation presented in \cite{alpaca}, we notice that it is necessary to eliminate the terms in $a^k a^{\dagger l}$ where $k \neq l$, a result that is obtained by applying the RWA approximation (\textit{Rotating-Wave Approximation}, which means neglecting the terms oscillating too quickly, i.e. the terms whose phase components do not cancel each other out); indeed, the \textsc{Ehrenfest} theorem applied to $a$ and $a^{\dagger}$ shows that $a^k a^{\dagger l} \ \propto \ e^{i w_r(l-k)t}$. We finally end up with the expression ($K$ is called \textsc{Kerr}'s constant):

$$\mathcal{H} = \hbar(w_r a^{\dagger}a + \frac{K}{2}(a^{\dagger}a)^2)$$

\newpage

By summing such an expression on the number of oscillators composing the network (which allows to obtain a system approaching the \textit{cat states}, opposite states that can be realized simultaneously, such as spins) and by adding to this model the contribution of the pump $p(t)$, we finally understand physically the origin of a first part of the Hamiltonian defined in \cite{alpaca}. \\

\subsubsection{Magnetic energy and quantum tunneling}

Most of the other terms, appearing in a double sum, are a consequence of the potential interaction between the spins (quantum tunneling is possible with, for instance, a spin disappearing at position $i$ and appearing at position $j$ or the other way round, leading to an operator $a_i^{\dagger}a_j + a_j^{\dagger}a_i$ with $[a_i,a_j^{\dagger}] = \delta_{i,j}$):

$$\mathcal{H}_T = -T(a_i^{\dagger}a_j + a_j^{\dagger} a_i)$$ where $T$ is the magnitude of the \textit{quantum tunneling effect} of the barrier separating two spins. \\

Finally, the \textsc{Ising} model assumes the existence of an external magnetic field $h$ interacting with the lattice of spins. Classically, the electrostatic potential energy is defined as the work of the magnetic force on the vector of the magnetic dipole moment $- \mu_S \cdot h$, and the observable magnetic moment of a spin is defined as $\mu_S = \frac{gQ}{2m}s$ with $g$ the constant \textit{g-factor}, $Q$ the charge, $m$ the mass and $s$ the spin angular momentum. Using the definition of the creation and annihilation operators defined in the previous subsection, we have $a_i + a_i^{\dagger} \ \propto \ -Q$, leading to a comprehensive Hamiltonian written as follows for a single spin ($\tilde{K}$ is a constant):

$$\mathcal{H}_M = \tilde{K}h_i(a_i + a_i^{\dagger})$$

\subsection{Hamiltonian of the quantum system}

In this section, we will focus on the actual resolution of Ising-type computational problems by minimizing adiabatically the energy of a network of \textsc{Kerr}-nonlinear parametric oscillators. The study carried out in the previous section justifies the following expression of the global Hamiltonian (observable as self-adjoint):
$$H(t)=\hbar\sum_{i=1}^{N\alpha}\left[\frac{K}{4}a_i^{\dagger 2}a_i^2-\frac{p(t)}{2}(a_i^{\dagger 2}+a_i^2)+\Delta_ia_i^\dagger a_i\right]-\hbar\xi_0\sum_{i=1}^{N\alpha}\sum_{j=1}^{N\alpha}J_{i,j}a_i^\dagger a_j+2\hbar A(t)\xi_0\sum_{i=1}^{N\alpha} h_i(a_i^\dagger+a_i)$$
where:
\begin{itemize}
    \item $N\alpha$ is the number of oscillators (N assets and weights encoded on $\alpha$ bits);
    \item $\hbar$ is the reduced \textsc{Planck} constant;
    \item $a_i$ (resp. $a_i^\dagger$) is the annihilation (resp. creation) operator;
    \item $K$ is \textsc{Kerr}'s constant;
    \item $p(t)$ is the pumping amplitude;
    \item $\Delta_i$ is the positive difference between the resonance frequency of oscillator $i$ and half the pumping frequency;
    \item $A(t)$ is a dimensionless positive time-dependent parameter;
    \item $\xi_0$ is a positive constant (dimension of a frequency).
\end{itemize}

Thereafter, we will not take into account the reduced \textsc{Planck} constant as it does not add anything to the financial model as long as we make sure the right units are being used.

\subsubsection{Approximate resolution} 

It is noteworthy that the output of the creation and annihilation operators might be complex whereas the output of the Hamiltonian is always real. Indeed, $a_{i}^{\dagger}a_{i} = ||a_i||^2 \in \mathbb{R}$ et $a_{i}^{\dagger 2} + a_{i}^{2} = (a_{i}^{\dagger} + a_{i})^{2} - 2||a_{i}||^{2} \in \mathbb{R}$.\\
If we define $\forall j \in \llbracket1,N\alpha\rrbracket, a_j = x_j+iy_j$ and $a_j^\dagger=x_j-iy_j$, the Hamiltonian can be expressed as follows:
$$H(\boldsymbol{x},\boldsymbol{y},t)=\sum_{i=1}^{N\alpha} \left [\frac{K}{4}(x_i^2+y_i^2)^2-\frac{p(t)}{2}(x_i^2-y_i^2)+\frac{\Delta_i}{2}(x_i^2+y_i^2)\right]-\frac{\xi_0}{2}\sum_{i=1,j=1}^{N\alpha}J_{i,j}(x_i x_j + y_i y_j)+2A(t)\xi_0\sum_{i=1}^{N\alpha} h_i x_i$$

This new expression of the Hamiltonian allows us to obtain the time derivatives of $x_i$ and $y_i$ (formulas deriving from Hamiltonian mechanics):

$$\dot{x_i}=\frac{\partial H}{\partial y_i}=\left[K(x_i^2+y_i^2)+p(t)+\Delta_i\right]y_i-\xi_0\sum_{j=1}^{N\alpha} J_{i,j}y_j$$
$$\dot{y_i}=-\frac{\partial H}{\partial x_i}=-\left[K(x_i^2+y_i^2)-p(t)+\Delta_i\right]x_i+\xi_0\sum_{j=1}^{N\alpha} J_{i,j}x_j-2A(t)\xi_0 h_i$$

That being said, the two previous equations are not easy to compute numerically; additional assumptions need to be made to solve them.

\subsubsection{Simplifying assumptions}

\begin{enumerate}
    \item \textbf{Variation of $y_i$}
    
    We initialize $(x_i, y_i)$ to $(0,0)$ as $\forall t, H(0,0,t) = 0$. As we need to study quantum spin of which we can only measure time averages, using ergodic hypothesis become a necessity. However, the Hamiltonian described earlier is not ergodic at all energy levels, and we must therefore approximate the system by an ergodic one. This can be achieved by separating variables into two categories: \textit{fast variables} and \textit{slow variables}. The fundamental idea is to state that the latter evolve because of the former: this is consistent with the physical context as $y_i$, the magnetic flux, is directly related to $x_i$, the charge. Thus, by categorizing $y_i$ as a slow variable, we can assume its evolution is slow compared to that of $x_i$. Considering the initial conditions, we can therefore consider that $y_i \ll x_i$ because $y_i$ evolves slowly around 0.
    
    \item \textbf{Uniformity of frequencies}
    
    It will be assumed that all frequencies $\Delta_i$ are equal to the same frequency $\Delta$.
    
    \item \textbf{Definition of A(t)}
    
    The function $A(t)$ can also be approximated. Indeed, the basic principle is to follow the evolution of a specific unstable equilibrium point when the underlying potential is slightly modified. \
    In order to do so, we define $A(t) \approx \left\{
    \begin{array}{ll}
        0 & \mbox{if } p(t) \ll \Delta \\
        \sqrt{\frac{p(t)-\Delta}{K}} & \mbox{if } p(t) \gg \Delta
    \end{array}
\right.$.\\

Thus, $A(t)$ is increasing with $p(t)$, allowing us to control the convergence of the solution.

    \item \textbf{Energy potential}
    
    We define $V(\boldsymbol{x},t)=\sum_{i=1}^{N\alpha} \huge{[} \frac{K}{4} x_i^4 + \frac{\Delta-p(t)}{2} x_i^2\huge{]} -\frac{\xi_0}{2}\sum_{i=1}^{N\alpha}\sum_{j=1}^{N\alpha}J_{i,j}x_i x_j + 2A(t)\xi_0\sum_{i=1}^{N\alpha} h_i x_i$ the energy potential as a function of $x_i$ only. Thus, the Hamiltonian can be expressed as follows:

\vspace{-0.3cm}

\begin{equation*}
\begin{split}
H(\textbf{x},\textbf{y},t) &= V(\textbf{x},t) + \frac{p(t)+\Delta}{2}\sum\limits_{i=1}^{N\alpha} y_{i}^{2} + \frac{K}{2}\sum\limits_{i=1}^{N\alpha} x_{i}^{2} y_{i}^{2}- \frac{\xi_0}{2}\sum\limits_{i=1}^{N\alpha}\sum\limits_{j=1}^{N\alpha} J_{i,j} y_{i}y_{j}\\
&= V(\textbf{x},t) + \frac{p(t)+\Delta - \xi_0 J_{i,i}}{2}\sum\limits_{i=1}^{N\alpha}y_{i}^{2}-\frac{\xi_0}{2}\sum\limits_{i=1}^{N\alpha}\sum\limits_{j=1}^{N\alpha}J_{i,j}y_{i}y_{j} + \frac{K}{2}\sum\limits_{i=1}^{N\alpha}x_{i}^{2}y_{i}^{2}
\end{split}
\end{equation*}

By the first assumption, we have $y_{i}^{2} = o(x_{i}^{2})$ so $x_{i}^{2}y_{i}^{2}$ is negligible compared to the term of the potential proportional to $x_i^4$. It is also reasonable to consider that the fluxes of the different spins are uncorrelated, which makes it possible to neglect the double sum.

\item \textbf{Compact form:} the Hamiltonian can be further simplified into a more convenient version that, given all the assumptions made so far, can be expressed as follows ($\delta_i := p(t)+\Delta-\xi_0 J_{i,i}$):
$$H_{SB}(\boldsymbol{x},\boldsymbol{y},t)=\sum_{i=1}^{N\alpha} \frac{\delta_i}{2}y_i^2 + V(\boldsymbol{x},t)$$
\end{enumerate}

\subsection{Adiabatic process}

The determination of the optimal \textit{ground state} is based on the bifurcation principle, which consists in modifying the eigenvalues of a system, and thus moving its equilibrium points, by changing the parameters that define the system. \\

\begin{figure}[!h]
\centering
\includegraphics[scale=0.3]{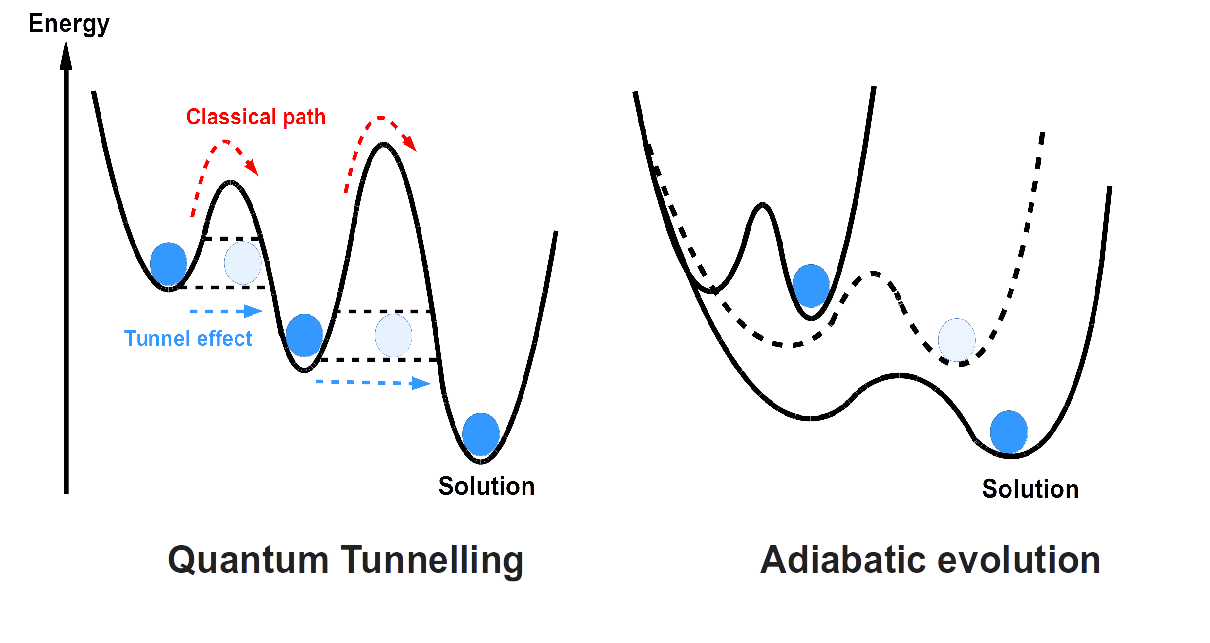}
\caption{Representation of an adiabatic process, illustration from \cite{medium}}
\end{figure}

In this specific situation, the parameters $p(t)$ and $A(t)$ will fulfill this role (actually only $p(t)$ as $A(t)$ evolves according to it). If the variation of these parameters is sufficiently slow (hence the \textit{adiabatic} nature), it is possible to track the evolution of the minimum from start to finish. The final values of $x_i$ and $y_i$ are therefore the approximate solution we are looking for.

\subsubsection{Definition of the variables}

With the new formulation of the Hamiltonian, it is possible to simplify the expression of the time derivatives of $x_i$ and $y_i$:

$$\dot{x_i}=\frac{\partial H_{SB}}{\partial y_i}=\delta_i y_i$$
$$\dot{y_i}=-\frac{\partial H_{SB}}{\partial x_i}=-\left[K x_i^2-p(t)+\Delta\right]x_i+\xi_0\sum_{j=1}^{N\alpha} J_{i,j}x_j - 2A(t)\xi_0 h_i$$

Solving these coupled differential equations leads to the values of $x_i$ and $y_i$ that makes it possible to determine the eigenstates of the Hamiltonian.

To set the parameters of the problem, we simply identify the constants of the physical system to those defined in the framework of our study. As $K$ and $\Delta$ are essentially involved in the formulation of $A(t)$, those constant's function is to regulate the evolution rate of the changing parameters. We can therefore choose their values so as to have complete control over $p(t)$ and $A(t)$. \\

To go back to our initial problem, we just have to recover the orientation of the spins by looking at the sign of the variables $x_i$. Indeed, $\forall i \in \llbracket 1,N \rrbracket$, $s_i=\mathrm{sign}(x_i)$ with $\mathrm{sign} : x \mapsto \left\{
    \begin{array}{ll}
        -1 & \mbox{if } x < 0 \\
        +1 & \mbox{if } x > 0 \\
        0 & \mbox{otherwise}
    \end{array}
\right.$. \\ We will consider that as long as $x_i$ is null, the state of equilibrium has not yet been reached. \\

With this modeling, we can determine the value of $s_i$ (then immediately the value of $b_i$) once we have converged towards the final equilibrium positions. Finally, the variables $b_i$ allows us to reconstruct the weights $w$: we have solved the problem of optimal asset allocation.

\subsubsection{Implementation via the \textsc{Euler} method}

In order to solve numerically these differential equations, we discretize time thanks to a step $\Delta_t$, and define the instants $t_n$ such that $t_n = n\Delta_t$ with $n\in\mathbb{N}$. The \textsc{Euler} problem therefore becomes:

$$x_i(t_{n+1})=x_i(t_n)+\Delta y_i(t_n)\Delta_t$$
$$y_i(t_{n+1})=y_i(t_n)-\left[ Kx_i^3(t_{n+1})+\left(\Delta - p(t_{n+1})\right)x_i(t_{n+1}) - \xi_0\sum_{j=1}^{N\alpha} J_{i,j}x_j(t_{n+1})+2A(t_{n+1})\xi_0 h_i\right]\Delta_t$$

However, in order to refine the accuracy of the \textsc{Euler} method and reduce the amount of heavy and time-consuming calculations, we introduce an integer parameter $M \geq 2$ that we use to obtain the symplectic formulation. The fundamental idea is computing a new \textsc{Euler} method between two consecutive time steps of the main one, only on low order values, to gain in precision while not increasing the time span of evolution. \\

Thus, for a given oscillator $(x_i,y_i)$ at a given time step $t_n$, we introduce new local variables $x_i^{(m)}$ and $y_i^{(m)}$, with $m \in \llbracket 0, M \rrbracket$, such that: 
$$x_i^{(0)}=x_i(t_n)$$ 
$$y_i^{(0)}=y_i(t_n)$$
$$x_i(t_{n+1})=x_i^{(M)}$$
$$y_i(t_{n+1}) = y_i^{(M)} + \xi_0 \left [ \sum\limits_{j=1}^{N\alpha}J_{i,j}x_j^{(M)} - 2A(t_{n+1})h_i\right ] \Delta_t$$
$$\forall m \in \llbracket 0, M-1 \rrbracket, \; x_i^{(m+1)}=x_i^{(m)} + \Delta y_i^{(m)} \delta_t$$
$$\forall m \in \llbracket 0, M-1 \rrbracket, \; y_i^{(m+1)}=y_i^{(m)} - \left [Kx_i^{(m+1)^3} + (\Delta - p(t_{n+1}))x_i^{(m+1)}\right ] \delta_t$$ with $\delta_t = \frac{\Delta_t}{M}$. \\

Hence, for each oscillator and at a given time step, the number of calculations required for $x$ are in $\mathcal{O}(M)$ and in $\mathcal{O}(M+N\alpha)$ for $y$. As the number of oscillators is equal to the number of assets $N$ multiplied by the number of bits of the integer representation of $w$, i.e. $\alpha$, we can conclude that the amount of calculations needed at each time step is in $\mathcal{O}(N\alpha(2M + N\alpha))$, leading to a global complexity of $\mathcal{O}(T(N\alpha)(2M + N\alpha))$ where $T$ is the total number of time steps used for the \textsc{Euler} method. Finally, we have to reconstruct the $w_i$ thanks to the $(\lim_{n \to + \infty} \mathrm{sign}(x_i(t_n)))_i$, which is an operation in $\mathcal{O}(N\alpha)$. \\

That being said, as the symplectic parameter $M$ is only supposed to take small integer values (usually between $2$ and $5$) in order not to make the calculations too cumbersome, $M$ can be neglected in the complexity. In summary, the complexity of the \textit{Simulated Bifurcation} algorithm is:

$$\mathcal{O} \left (TN^2\alpha^2 \right )$$ which is obviously more convenient than the $\mathcal{O}(2^{N\alpha})$ complexity of the brute-force algorithm.

\begin{figure}[!h]
\centering
\includegraphics[scale=0.4]{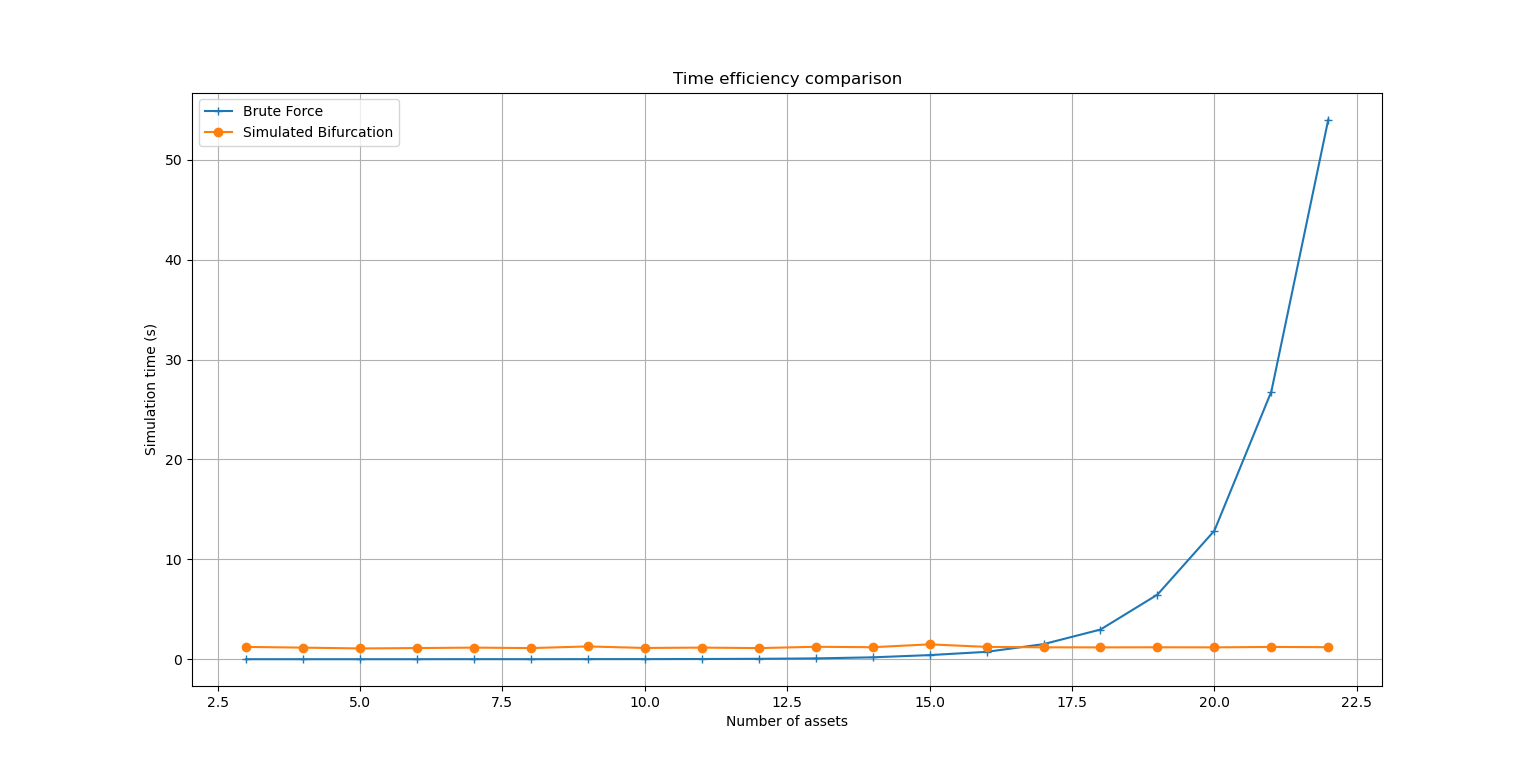}
\caption{Comparison of the time efficiency of \textit{Simulated Bifurcation} and naive brute-force ($\alpha=1$)}
\end{figure}

\subsubsection{Stop criterion}

At this point, an interesting question is how to stop the calculation of the virtually infinite \textsc{Euler} method. \\

The naive way to proceed would be to choose a number of steps $T$ and to stop the simulation once this number of steps is reached. Although this may be sufficient in some cases, it is not possible to guarantee the general convergence of the \textsc{Euler} method after these $T$ steps - in the sense that there is no guarantee that all spins will have effectively bifurcated and reached their \textit{steady state}. On the other hand, there is no explicit formulation linking a given definition of convergence to a number of steps. \\

A more refined method consists in sampling the spins over time and compare the values taken by each of these at different times, using a window of fixed size. To illustrate this method more explicitly, let us introduce the following quantities:

\begin{itemize}
    \item $T_S$: the \textbf{sampling period}, i.e. the number of steps one should wait before sampling the value of the spins;
    \item $n$: the \textbf{window's size}, i.e. the number of samples kept in memory in the window and which will be compared with each other.
\end{itemize}

Formally, the time window is a matrix of size $N\alpha \times n$ that contains the values of the spins in the last $n$ samples. Every $T_S$ steps in the \textsc{Euler} method, all columns of the window are shifted one rank to the left and the last column is replaced by the current spin vector. Denoting $W_t^S$ the window value and $s_t$ the spin vector at the step $t$, two cases are to be treated:

\begin{itemize}
    \item either $t \equiv 0 \; [T_S]$ and then $\forall i \in \llbracket 1, n-1 \rrbracket, \left[W_{t+1}^S\right]_i = \left[W_{t}^S\right]_{i+1}$ and $\left[W_{t+1}^S\right]_n = s_{t+1}$
    \item either $t \nequiv 0 \; [T_S]$ and then $W_{t+1}^S = W_t^S$
\end{itemize}

\textbf{Remark: }Note that, by convention, we have $W_0^S = 0_{\mathcal{M}_{N\alpha,n}(\mathbb{R})}$.\\

As previously stated, we want the \textsc{Euler} method to stop if the spins have all bifurcated, i.e. their sign is constant over the last $n$ sampled steps. Mathematically, this can be translated into the nullity of the variance of each of the rows in the window $W_t^S$. Thus at each step, after having (if necessary) updated the value of the window and having made sure that no column of zeros remains (which would mean that we have sampled less than $n$ times the spins during the Euler method), the calculation of the variance of each of its rows is all one needs to perform. If all of them are zero, then the spins have all bifurcated and the calculation can stop there. \\

A discussion is necessary when choosing the parameters $T_S$ and $n$. Indeed, their values will greatly influence the decision of whether to continue or not the method and play a major role in the total computation time of the \textsc{Euler} method. \\

To begin, let us deal with the choice of the sample period. It is immediately obvious that the smaller the sampling period, the more the spin samples will be temporally correlated, and vice versa. A period that is too small would only have a local effect and could bias the result because nothing \textit{a priori} prevents the spins from changing sign during the simulation. Indeed, such a change could occur after a period of stability and would then go unnoticed. On the contrary, a large period ensures that spins will be relatively decorrelated and an equality of these samples sufficiently spaced in time gives more credibility to the hypothesis of permanent and stable bifurcation. However, a large sampling period also means waiting more steps before sampling the spins and thus letting the algorithm run longer. \\

With respect to the size of the window, a similar analysis is required. A small window size ensures short computation times but only yields an analysis of the local evolution of the spins, which induces the same problem obtained in the case of a small sampling period. Similarly, a larger window will significantly reduce the error margin at the cost of a non-negligible increase in computation time. \\

In terms of temporal complexity, column shifts and variance calculations are in $O(N\alpha \times n)$, i.e. the window's dimensions. These calculations take place every $T_S$ steps, each step costing $O\left((N\alpha)^2\right)$. This leads to a total complexity of the order of $O\left(T_S(N\alpha)^2 + nN\alpha\right)$ (taking as a reasonable assumption that the \textsc{Euler} method will always converge in a finite time). In view of this complexity, one would want to pick small values for $T_S$ and $n$ but one should also keep in mind that the final accuracy is inversely related to them - and that accuracy and computation time are both of major importance in this situation.

\subsubsection{Hamiltonian parameters}

Parameters relative to the number of assets, the available funds or the risk coefficient may vary at the user's convenience. However, the Hamiltonian parameters must be set once and for all in a way that leads to satisfying results. Hereafter, we will use the following values of these very parameters (most of them are established in \cite{combinatorial}): 

\begin{itemize}
    \item $K = 1$
    \item $\Delta = 1$
    \item $\xi_0 = \frac{0.7\Delta}{\sigma \sqrt{N\alpha}}$ where $\sigma$ is the standard deviation of the terms of the $J$ matrix
    \item $p : t \mapsto 0.01t$, $\forall t \geq 0$
\end{itemize}

\vspace{0.2cm}

\section{\textsc{Python} implementation}

To give more practical insights into the topic of portfolio optimization, we will implement in \textsc{Python} the algorithm described previously. Due to the exponential time complexity of the brute-force verification, we will only check the validity of the algorithm on small sets of financial instruments ($\leq 10$), and provide a statistical study underlying the quality of the obtained approximation.

\subsection{Datasets}

\subsubsection{Test dataset}

When we need to explicitly verify the quality of the obtained approximation (by resorting to brute-force algorithms that require a dataset of reasonable size), we will use the following dataset \cite{bnp}, where $\mu$ are the expected returns, $\sigma$ the volatilities and $\Sigma$ the covariance matrix:

\begin{equation*}
\begin{split}
    \mu &= [0.06,0.07,0.095,0.015,0.013,0.032,0]^T \\
    \sigma &= [0.143,0.174,0.212,0.043,0.040,0.084,0.005]^T \\
    \Sigma &= \begin{bmatrix} 2.04\times 10^{-2} & 2.04\times 10^{-2} & 2.36\times 10^{-2} & 6.14\times 10^{-4} & 0 & 6.00\times 10^{-3} & 0 \\
 2.04\times 10^{-2} & 3.02\times 10^{-2} & 3.13\times 10^{-2} & 8.97\times 10^{-4} & 5.56\times 10^{-4} &
  9.20\times 10^{-3} & 0 \\
 2.36\times 10^{-2} & 3.13\times 10^{-2} & 4.49\times 10^{-2} & 4.55\times 10^{-4} & 2.54\times 10^{-4} &
  1.26\times 10^{-2} & 0 \\
 6.14\times 10^{-4} & 8.97\times 10^{-4} & 4.55\times 10^{-4} & 1.84\times 10^{-3} & 1.11\times 10^{-3} &
  7.22\times 10^{-4} & 0 \\
 0 & 5.56\times 10^{-4} & 2.54\times 10^{-4} & 1.11\times 10^{-3} & 1.60\times 10^{-3} &
  7.72\times 10^{-4} & 0 \\
 6.00\times 10^{-3} & 9.20\times 10^{-3} & 1.26\times 10^{-2} & 7.22\times 10^{-4} & 7.72\times 10^{-4} &
  7.05\times 10^{-3} & 0 \\
 0 & 0 & 0 & 0 & 0 & 0 & 2.5\times 10^{-5}\end{bmatrix}
\end{split}
\end{equation*}

The assets represent the following sequence:

\begin{enumerate}
    \item American equities;
    \item European equities;
    \item Emerging market equities;
    \item European sovereign bonds;
    \item American sovereign bonds;
    \item Emerging market bonds;
    \item Sovereign letters of credit and cash.
\end{enumerate}

\subsubsection{Complete dataset}

As an illustration of the practical usefulness of the \textit{Simulated Bifurcation} algorithm, we have extracted historical data from \textsc{Yahoo! Finance} (closing prices of 441 assets belonging to the S\&P500 index), that was processed during the 02/2003 — 02/2021 period on the \textsc{New York Stock Exchange}. More specifically, we have calculated the daily returns and used such \textit{dataframe} to estimate the covariance matrix; full details are given in the annex. \\

Obviously, dealing with such a large amount of information would be impractical in a standard context, whereas we are able to obtain usable approximations using laptop processors (\textsc{Apple} M1 ARM SoC) in a few minutes. Furthermore, the process could be hastened by converting the \textsc{Python} code into C or even into fast machine language with tools such as \textsc{Numba}, paving the way for high-frequency applications (forex triangular arbitrage for instance).

\subsection{The choice of $\alpha$}

As presented previously, we must choose $\alpha\in\mathbb{N}$ such that $\forall t, \forall i, w_{it} \leq 2^{\alpha}-1$ and striking a balance between accuracy and computation time of the algorithm. To get insights into the evolution of the practical accuracy of the algorithm as a function of $\alpha$, we will present here the results of a statistical study aiming at analyzing the quality of the approximations provided by \textit{Simulated Bifurcation}. \\

More specifically, we will compare the optimal result obtained with a brute-force algorithm and the solution yielded by our program for different values of $\alpha$ and different number of assets (the time complexity of the brute-force being in $\mathcal{O}(2^{N\alpha})$, we will be forced to limit our exhaustive verification to $\alpha \leq 4$ and $N \leq 5$). \\

For each couple $(\alpha,N)$, we ran $50$ simulations using random symmetric positive matrices $\Sigma$ (that will represent covariance matrices, with coefficients in the order of magnitude of $10^{-4}$) and return vectors $\mu$ (with coefficients in the order of magnitude of $10^{-2}$) and compared the average results.

\subsubsection{Exact accuracy}

This first table shows the percentage of the simulations which led the \textit{Simulated Bifurcation} algorithm to yield the exact same result as the one given by the brute-force algorithm ($\times$ means that the couple (assets, bits) was computationally intractable and was therefore not pursued):

\begin{center}
    \begin{tabular}{ |c|c|c|c|c|c|c|c|c|c|c|c|c|c| } 
        \hline
        \textbf{↓ bits ($\alpha$)} / \textbf{assets →} & \textbf{2} & \textbf{3} & \textbf{4} & \textbf{5} & \textbf{6} & \textbf{7} & \textbf{8} & \textbf{9} & \textbf{10} & \textbf{11} & \textbf{12} & \textbf{13} & \textbf{14} \\
        \hline
        \textbf{1} & 100 & 100 & 96 & 98 & 98 & 98 & 98 & 97 & 99 & 94 & 95 & 88 & 89 \\ 
        \hline
        \textbf{2} & 99 & 98 & 99 & 100 & 97 & 93 & $\times$ & $\times$ & $\times$ & $\times$ & $\times$ & $\times$ & $\times$ \\ 
        \hline
        \textbf{3} & 98 & 97 & 89 & $\times$ & $\times$ & $\times$ & $\times$ & $\times$ & $\times$ & $\times$ & $\times$ & $\times$ & $\times$ \\ 
        \hline
        \textbf{4} & 71 & 65 & $\times$ & $\times$ & $\times$ & $\times$ & $\times$ & $\times$ & $\times$ & $\times$ & $\times$ & $\times$ & $\times$ \\ 
        \hline
        \textbf{5} & 29 & $\times$ & $\times$ & $\times$ & $\times$ & $\times$ & $\times$ & $\times$ & $\times$ & $\times$ & $\times$ & $\times$ & $\times$\\ 
        \hline
        \textbf{6} & 9 & $\times$ & $\times$ & $\times$ & $\times$ & $\times$ & $\times$ & $\times$ & $\times$ & $\times$ & $\times$ & $\times$ & $\times$ \\ 
        \hline
        \textbf{7} & 4 & $\times$ & $\times$ & $\times$ & $\times$ & $\times$ & $\times$ & $\times$ & $\times$ & $\times$ & $\times$ & $\times$ & $\times$ \\ 
        \hline
    \end{tabular}
\end{center}

\subsubsection{Relative gap}

The following two tables present for each pair (assets, bits) the average relative deviation ($\times 10^{-4}$) over 100 simulations between the optimal solution and the approximation obtained thanks to the \textit{Simulated Bifurcation} algorithm. The first one concerns the Ising energy and the second the value of the Markowitz utility function. The difference between the two tables comes from the fact that the reconstruction of an integer vector from a vector of spins induces additional errors due to the weights associated to each spin:

\begin{center}
    \begin{tabular}{ |c|c|c|c|c|c|c|c|c|c|c|c|c|c| } 
        \hline
        \textbf{↓ bits} / \textbf{assets →} & \textbf{2} & \textbf{3} & \textbf{4} & \textbf{5} & \textbf{6} & \textbf{7} & \textbf{8} & \textbf{9} & \textbf{10} & \textbf{11} & \textbf{12} & \textbf{13} & \textbf{14} \\
        \hline
        \textbf{1} & 0 & 0 & 3.92 & 0 & 2.75 & 0.65 & 0.40 & 2.61 & 0.22 & 2.33 & 3.87 & 7.11 & 6.12 \\ 
        \hline
        \textbf{2} & 1.68 & 0.44 & 0.68 & 0 & 0.24 & 4.60 & $\times$ & $\times$ & $\times$ & $\times$ & $\times$ & $\times$ & $\times$ \\ 
        \hline
        \textbf{3} & 3.13 & 3.78 & 5.52 & $\times$ & $\times$ & $\times$ & $\times$ & $\times$ & $\times$ & $\times$ & $\times$ & $\times$ & $\times$ \\ 
        \hline
        \textbf{4} & 79.37 & 135.35 & $\times$ & $\times$ & $\times$ & $\times$ & $\times$ & $\times$ & $\times$ & $\times$ & $\times$ & $\times$ & $\times$ \\ 
        \hline
        \textbf{5} & 1625.38 & $\times$ & $\times$ & $\times$ & $\times$ & $\times$ & $\times$ & $\times$ & $\times$ & $\times$ & $\times$ & $\times$ & $\times$\\ 
        \hline
        \textbf{6} & 6737.86 & $\times$ & $\times$ & $\times$ & $\times$ & $\times$ & $\times$ & $\times$ & $\times$ & $\times$ & $\times$ & $\times$ & $\times$ \\ 
        \hline
        \textbf{7} & 5129.21 & $\times$ & $\times$ & $\times$ & $\times$ & $\times$ & $\times$ & $\times$ & $\times$ & $\times$ & $\times$ & $\times$ & $\times$ \\ 
        \hline
    \end{tabular}
\end{center}

\begin{center}
    \begin{tabular}{ |c|c|c|c|c|c|c|c|c|c|c|c|c|c| } 
        \hline
        \textbf{↓ bits} / \textbf{assets →} & \textbf{2} & \textbf{3} & \textbf{4} & \textbf{5} & \textbf{6} & \textbf{7} & \textbf{8} & \textbf{9} & \textbf{10} & \textbf{11} & \textbf{12} & \textbf{13} & \textbf{14} \\
        \hline
        \textbf{1} & 0 & 0 & 1.96 & 0 & 1.37 & 0.32 & 0.19 & 1.28 & 0.11 & 1.16 & 2.09 & 3.75 & 2.99 \\ 
        \hline
        \textbf{2} & 0.82 & 0.21 & 0.31 & 0 & 0.12 & 2.16 & $\times$ & $\times$ & $\times$ & $\times$ & $\times$ & $\times$ & $\times$ \\ 
        \hline
        \textbf{3} & 101.31 & 2.11 & 2.40 & $\times$ & $\times$ & $\times$ & $\times$ & $\times$ & $\times$ & $\times$ & $\times$ & $\times$ & $\times$ \\ 
        \hline
        \textbf{4} & 60.41 & 49.99 & $\times$ & $\times$ & $\times$ & $\times$ & $\times$ & $\times$ & $\times$ & $\times$ & $\times$ & $\times$ & $\times$ \\ 
        \hline
        \textbf{5} & 548.72 & $\times$ & $\times$ & $\times$ & $\times$ & $\times$ & $\times$ & $\times$ & $\times$ & $\times$ & $\times$ & $\times$ & $\times$\\ 
        \hline
        \textbf{6} & 1505.53 & $\times$ & $\times$ & $\times$ & $\times$ & $\times$ & $\times$ & $\times$ & $\times$ & $\times$ & $\times$ & $\times$ & $\times$ \\ 
        \hline
        \textbf{7} & 6423.33 & $\times$ & $\times$ & $\times$ & $\times$ & $\times$ & $\times$ & $\times$ & $\times$ & $\times$ & $\times$ & $\times$ & $\times$ \\ 
        \hline
    \end{tabular}
\end{center}

\vspace{0.5cm}

\section{Particular case: one-bit weights}

A particularly interesting subproblem one can solve using the same methods described previously is the one in which we only consider one-bit weights: this is equivalent to deciding whether or not we should invest in a particular asset. \\

All in all, we are looking for the \textbf{optimal subset of assets}. There are two main benefits to such a consideration: first of all, it provides small size investors with the best allocation of their money (all the more significant for stocks and options, since it is not possible to purchase fractions of stocks without resorting to CFDs). Furthermore, cherrypicking the best assets is valuable for shortening the execution time of every optimization algorithm, especially since the complexities are at least polynomial in the number of assets.

\subsection{Optimal subset of assets}

For the same reasons one cannot prove that the \textit{Simulated Bifurcation} algorithm provides a good approximation for a large number $n$ of assets, it is not feasible to explicitly check the output obtained with each element of $\{0,1\}^N$. We therefore only provide here a brute-force verification for $N=7$ and the pair $(\mu,\Sigma)$ defined earlier in this paper (we will look for $\underset{w \in \{0,1\}^N}{\mathrm{argmax}} \ \Phi(w)$ with $\Phi(w) = w^{T}\mu - \frac{\gamma}{2} w^{T}\Sigma w$).

\begin{itemize}
    \item We generate all binary representations of integers between $0$ and $2^N-1$;
    \item We front-fill the elements with zeros such that list $[0]$ becomes $[0,...,0]^T$;
    \item We compute the function $\Phi(w)$ for each element and extract the maximum.
\end{itemize}

As expected, the maximum risk-aware return matches the output of the algorithm ($\simeq 0.131$ in this case).

\subsection{Intuitive behavior with respect to returns}

It would be reasonable to expect that the optimum would only select the assets that provide non-negative returns. Indeed, even if an asset has an important volatility and a slightly negative return (which would make it possible for the actual return to be positive), it would not be coherent to select it as the expected return with such an asset would decrease.

To illustrate this property, we set:

\begin{equation*}
\begin{split}
    \mu &= [0.06,0.07,0.095,0.015,0.013,0.032,0.01]^T\\
    \text{instead of } \mu &= [0.06,0.07,0.095,0.015,0.013,0.032,0]^T
\end{split}
\end{equation*}

In other words, we suppose that cash has a very slightly non-zero return in this scenario (deposit in a savings account). As expected, the \textit{Simulated Bifurcation} algorithm now also selects the asset \textit{cash}. If we consider $-\mu$ instead (only negative returns), the algorithm obviously returns a null vector.

\subsection{Intuitive behavior with respect to volatility}

Similarly, one would expect the optimal solution to minimize the uncertainty range of returns, which translates into selecting only assets with low volatility if returns are the same. In order to verify such a behavior, we set $\mu = 0.01 \times \mathds{1}$ and we observe that the optimal subset discards assets with a relatively high volatility:

$$w_{opt} = [0,0,0,1,1,1,1]^T$$ because $\sigma = [0.143,0.174,0.212,0.043,0.040,0.084,0.005]^T$.

\subsection{Results}

Now that we have a broader understanding of the one-bit case, we will apply the method to the complete dataset described in a previous subsection. An intuitive \textsc{Plotly} interface allowing for a clear visualization of the simulations is implemented in the Appendix.

\begin{figure}[!h]
\centering
\includegraphics[scale=0.4]{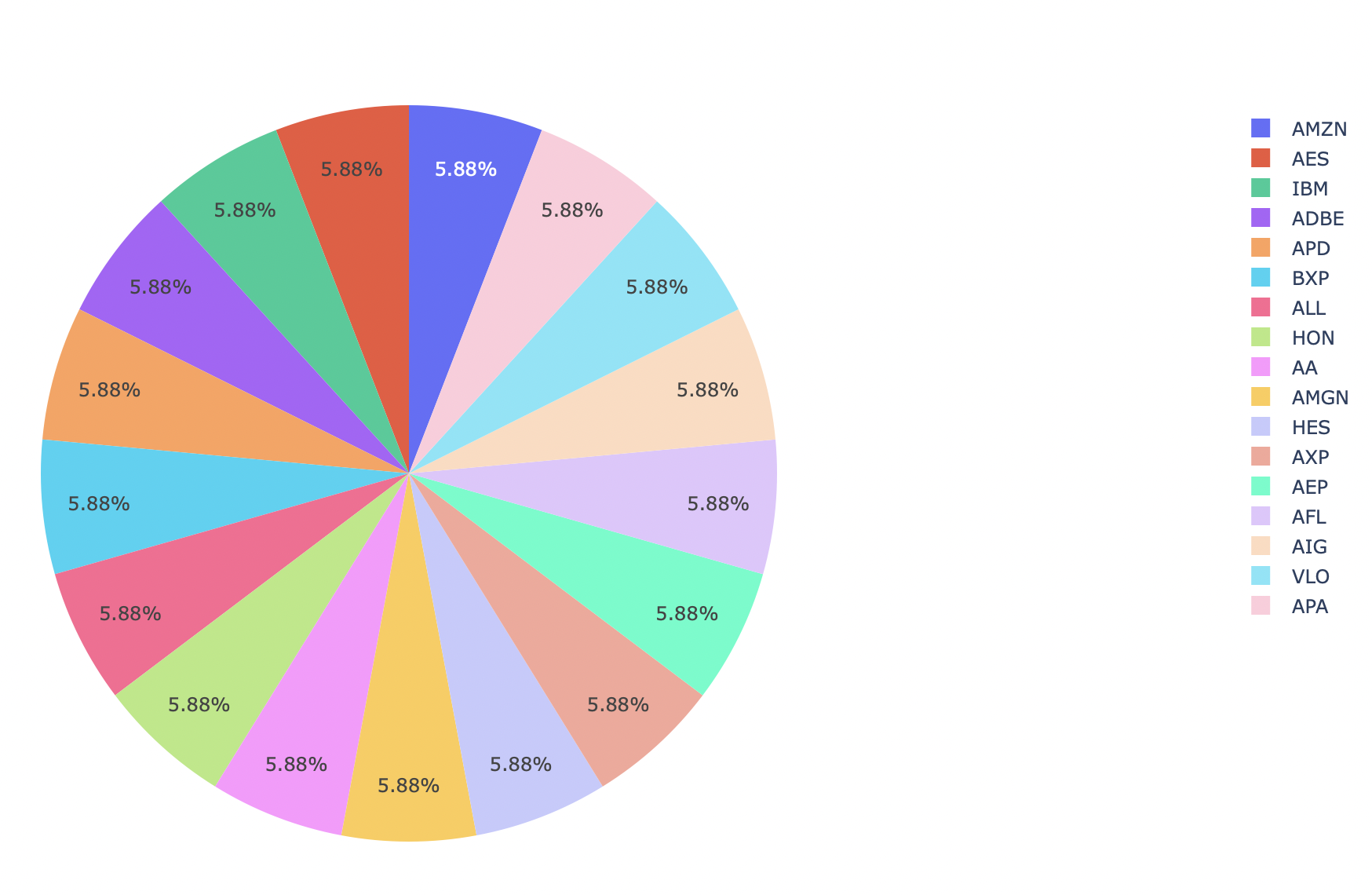}
\caption{Computed optimal subset of the first 20 assets (lexicographic order) for the 01/03/2021}
\end{figure}

\hspace{0.2cm}

First of all, it is important to specify that the algorithm runs in only 5 seconds, a remarkable performance since we are using a dataset containing 441 assets. As regards the allocation themselves, one can notice that the obtained subsets are of smaller size compared to the original one (about 120 assets are chosen out of the 441 being considered), somewhat meeting the desired synthetic nature. \\

At this point, it makes sense to compare the performance $\Phi(w)$ of the suggested portfolio with the one obtained in the case of an equally-weighted allocation ($ew$) on the complete asset class. We obtain for the 01/03/2021 that:

\begin{equation*}
\begin{split}
    \Phi(w_{ew}) &\simeq -10 \\
    \Phi(w_{sb}) &\simeq 2.82
\end{split}
\end{equation*} and this clearly shows that the $w_{sb}$ allocation is a much better risk-aware one. \\

While it is impossible to provide a mathematical proof of optimality (a brute-force enumeration would be unfeasible at this scale), the obtained behavior seems intuitive. For instance, \textsc{Amazon} is selected in the computed allocation for the 9th of July 2020, because the underlying financial asset has surged in July 2020. Moreover, the algorithm respects the principle of diversification since most of the selected assets belong to different sectors. \\

In the light of such promising results, we decided to analyze the performance of the \textit{Simulated Bifurcation} algorithm on subsets of fixed cardinality (ranging from $6$ to $18$ assets) of the complete dataset. More specifically, the graph below presents the average ratio of weights matching the optimal ones obtained thanks to almost $2000$ simulations ($150$ simulations per point).

\begin{figure}[H]
\centering
\includegraphics[scale=0.7]{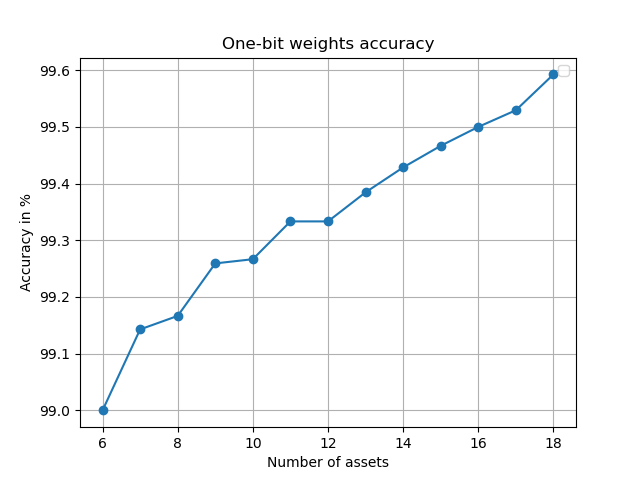}
\caption{Accuracy of the \textit{Simulated Bifurcation} algorithm ($\alpha=1$ case)}
\end{figure}

It is also noteworthy that the optimal allocation was generally returned by the algorithm $138$ times out of the $150$ simulations, and in the other cases the approximation usually only differed by a single bit. \\

Unfortunately, it is technically impossible to pursue this comparison for far greater numbers of assets because of the combinatorial explosion of the brute-force enumeration. Yet, these partial results could indicate that the method is particularly interesting when $\alpha=1$ (quadratic time complexity with respect to $N$ and at least 99\% accuracy).

\newpage

\section{Conclusion}

The contribution of this paper is to explore the \textit{Simulated Bifurcation} algorithm, both from theoretical and practical standpoints. First of all, we have explained the physical foundations of the method, derived from the minimization of the energy of a network of \textsc{Kerr}-nonlinear parametric oscillators. In practical terms, we have implemented the algorithm in \textsc{Python} and yielded insightful results (90.4\% \textsc{Hamming} accuracy for 4 assets encoded on 5 bits, and even 99\% accuracy on average with one-bit weighting) from 441 assets belonging to the S\&P500 index in less than 5 seconds (and reasonable execution times for larger bit encodings). \\

\begin{figure}[!h]
\centering
\includegraphics[scale=0.55]{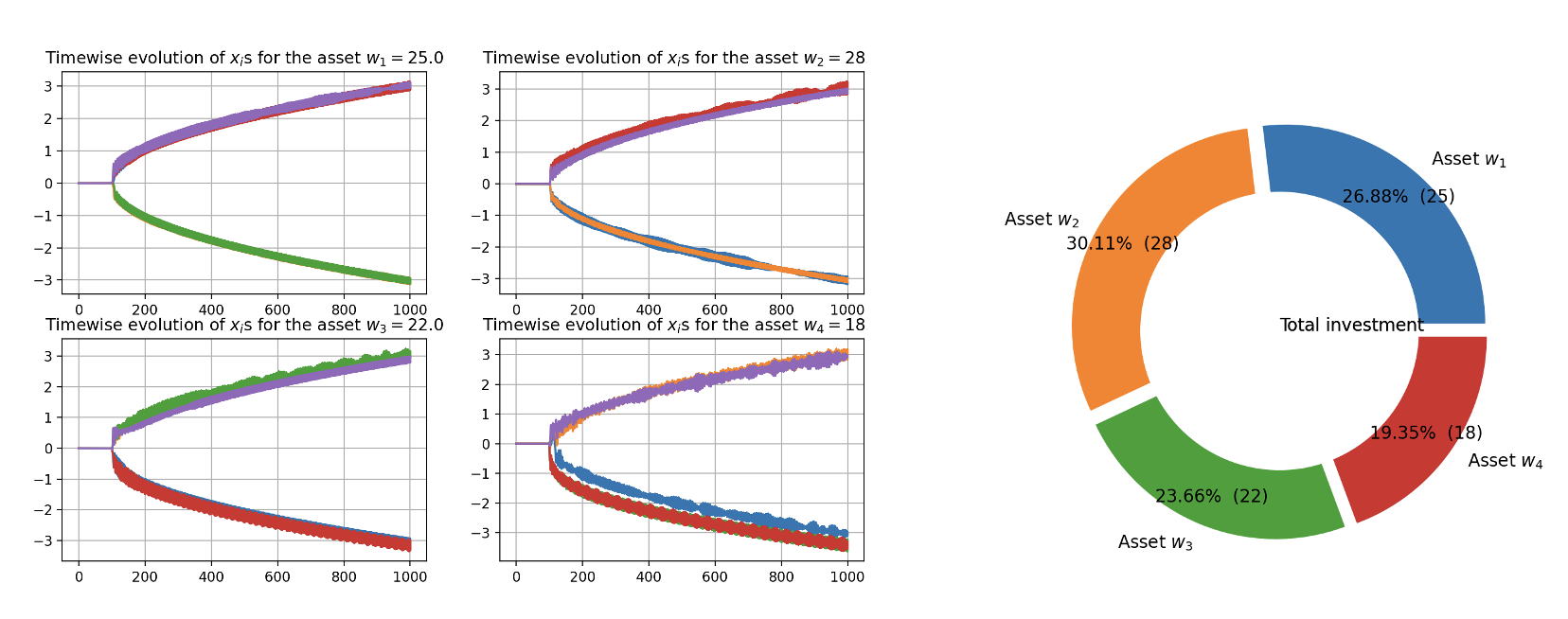}
\caption{Example of a computed optimal allocation of 4 assets}
\end{figure}

Moreover, we have focused on the particular case of binary allocations allowing the user to obtain the optimal equally weighted sub-portfolio. The financial benefit of this result has been discussed in the dedicated section and the possibility of obtaining such a solution in a few seconds on a personal computer paves the way for high-frequency applications. \\

Finally, it is worth mentioning that many NP problems can be expressed thanks to an \textsc{Ising} formulation (such as \textit{Knapsack}, \textit{Partitioning problems}, \textit{Steiner Trees} and many others described in \cite{NP formulations}), enabling an efficient resolution thanks to \textit{Simulated Bifurcation}.

\newpage

\section{Appendix}

\addtocontents{toc}{\protect\setcounter{tocdepth}{1}}

\subsection{Exhaustive dataset}

As an illustration of the practical usefulness of the \textit{Simulated Bifurcation} algorithm, we have extracted the closing prices data of the following 441 assets belonging to the S\&P500 index from \textsc{Yahoo! Finance} (sorted by lexicographical order): \\

A, AA, AAP, AAPL, ABC, ABMD, ACN, ADBE, ADI, ADM, ADP, ADS, ADSK, AEE, AEP, AES, AFL, AIG, AIV, AJG, AKAM, ALB, ALGN, ALL, ALXN, AMAT, AMD, AME, AMG, AMGN, AMT, AMZN, AN, ANSS, ANTM, AON, APA, APD, APH, ARE, ASH, ATI, ATVI, AVB, AXP, AZO, BA, BAC, BAX, BBBY, BBY, BC, BDX, BEN, BIG, BIIB, BK, BKNG, BKR, BLK, BLL, BMY, BSX, BWA, BXP, C, CAG, CAH, CAR, CAT, CB, CCEP, CCI, CCL, CDNS, CERN, CHD, CHRW, CI, CIEN, CINF, CL, CLF, CLX, CMA, CMCSA, CME, CMS, CNC, CNP, CNX, COF, COG, COO, COP, COST, CPB, CPRT, CSCO, CSX, CTAS, CTB, CTSH, CTXS, CVS, CVX, D, DD, DDS, DE, DFODQ, DGX, DHI, DHR, DIS, DISH, DLTR, DLX, DOFSQ, DRI, DTE, DUK, DVA, DVN, DXC, EA, EBAY, ECL, ED, EFX, EHC, EIX, EL, EMN, EMR, ENDP, EOG, EQIX, EQR, EQT, ES, ESS, ETN, ETR, EVRG, EW, EXC, EXPD, F, FCX, FDX, FE, FFIV, FHI, FHN, FIS, FISV, FITB, FLIR, FLR, FLS, FMC, FMCC, FNMA, FTI, FTRCQ, GD, GE, GILD, GIS, GL, GLW, GME, GPC, GPN, GPS, GS, GWW, HAL, HAS, HBAN, HD, HES, HIG, HOG, HOLX, HON, HP, HPQ, HRB, HRL, HSIC, HSY, HUM, IBM, IDXX, IFF, ILMN, INCY, INTC, INTU, IP, IPG, IRM, ISRG, ITT, ITW, IVZ, J, JBL, JCI, JEF, JNJ, JNPR, JPM, JWN, K, KBH, KEY, KIM, KLAC, KMB, KMX, KO, KR, KSS, KSU, L, LB, LEG, LEN, LH, LHX, LIN, LLY, LMT, LNC, LOW, LPX, LRCX, LUV, M, MAC, MAR, MAS, MAT, MBI, MCD, MCHP, MCK, MCO, MDLZ, MDP, MDT, MET, MGM, MHK, MKC, MLM, MMC, MMM, MNST, MO, MOS, MRK, MRO, MS, MSFT, MSI, MTB, MTD, MTG, MTW, MU, MXIM, NAV, NBR, NCR, NDAQ, NEE, NFLX, NI, NKE, NKTR, NLOK, NOC, NOV, NSC, NTAP, NTRS, NUE, NVDA, NWL, NYT, O, ODFL, ODP, OI, OKE, ORCL, ORLY, OXY, PAYX, PBCT, PBI, PCAR, PCG, PDCO, PEAK, PEG, PEP, PFE, PFG, PG, PGR, PH, PHM, PKI, PLD, PNC, PNR, PNW, PPG, PPL, PRGO, PRU, PSA, PVH, PWR, PXD, QCOM, R, RCL, REGN, RF, RHI, RIG, RL, RMD, ROK, ROP, ROST, RRC, RRD, RSG, RTX, SANM, SBAC, SBUX, SCHW, SEE, SHW, SIG, SITC, SIVB, SJM, SLB, SLM, SNA, SNPS, SNV, SO, SPG, SPGI, SRCL, SRE, SSP, STT, STX, STZ, SWK, SWKS, SWN, SYK, SYY, T, TAP, TER, TEX, TFC, TFX, TGNA, TGT, THC, TJX, TMO, TPR, TROW, TRV, TSCO, TSN, TT, TTWO, TXN, TXT, UHS, UIS, UNH, UNM, UNP, UPS, URI, USB, VALPQ, VAR, VFC, VIAV, VLO, VMC, VNO, VRSN, VRTX, VTR, VTRS, VZ, WAT, WBA, WDC, WEC, WELL, WFC, WHR, WLTW, WM, WMB, WMT, WOR, WST, WY, WYNN, X, XEC, XEL, XLNX, XOM, XRAY, XRX, YUM, ZBH, ZBRA, ZION

\vspace{20pt}

Regarding $\mu$ and $\Sigma$ used as an input for the algorithm, we have adopted the following definitions:

$$\mu_{i}(t) = \frac{p_{i,t} - p_{i,t-1}}{p_{i,t-1}}$$ i.e. the daily return for asset $i$ ($p_{i,t}$ being the closing price of asset $i$ on day $t$).

$$\Sigma = Cov((\mu(t))_{02/2003 \leq t \leq 02/2021})$$ with $\mu(t)=[\mu_{A}(t) \ \dots \ \mu_{ZION}(t)]$

\subsection{\textsc{Github} repository}

In this repository, we have implemented the \textit{Simulated Bifurcation} algorithm in \textsc{Python}. To ensure broad understanding of the codes, we have adopted the same naming conventions as the ones detailed in this paper.

\begin{center}
    \url{https://github.com/bqth29/simulated-bifurcation-algorithm}
\end{center}

\end{document}